\documentclass[prb,altaffillsymbol,superscriptaddress, 
citeautoscript,amsfonts,amssymb,amsmath,twocolumn,floatfix,altaffillsymbol]
{revtex4}

\usepackage{graphicx}
\begin{document}

\title{Magnetic studies of the lightly Ru doped perovskite rhodates Sr(Ru,Rh)O$_3$ }

\author{K. Yamaura}
\email[E-mail at:]{YAMAURA.Kazunari@nims.go.jp}
\homepage[Fax.:]{+81-29-860-4674}
\affiliation{Superconducting Materials Center, National Institute for Materials Science, 1-1 Namiki, Tsukuba, Ibaraki 305-0044, Japan}

\author{D.P. Young}
\affiliation{Department of Physics and Astronomy, Louisiana State University,
Baton Rouge, LA 70803}

\author{E. Takayama-Muromachi}
\affiliation{Superconducting Materials Center, National Institute for Materials
Science, 1-1 Namiki, Tsukuba, Ibaraki 305-0044, Japan}

\begin{abstract}
The solid solution between the ferromagnetic metal SrRuO$_3$ and the enhanced paramagnetic metal SrRhO$_3$ was recently reported [K. Yamaura et al., Phys. Rev. B 69 (2004) 024410], and an unexpected feature was found in the specific heat data at $x$=0.9 of SrRu$_{1-x}$Rh$_x$O$_3$.  The feature was reinvestigated further by characterizing additional samples with various Ru concentrations in the vicinity of $x$=0.9.  Specific heat and magnetic susceptibility data indicate that the feature reflects a peculiar magnetism of the doped perovskite, which appears only in the very narrow composition range $0.85$$<$$x$$\le$$0.95$.
\end{abstract}

\maketitle

Recently a full-range solid solution between the ferromagnetic metal SrRuO$_3$ ($T_{\rm c}$$\sim$160 K) and the enhanced paramagnetic metal SrRhO$_3$ was reported, and the 4$d$-correlated electrons in the perovskite-structure basis, ranging from 4$d^4$ to 4$d^5$ configurations, were investigated \cite{1}. The study was motivated by the fact that both the Ru and Rh perovskite-based families show intriguing correlated electronic features, including, unconventional superconductivity, itinerant ferromagnetism, and quantum critical behavior \cite{2}.  In the former study on the solid solution, the ferromagnetic state was gradually suppressed by the Rh substitution and vanished near $x$=0.6 in SrRu$_{1-x}$Rh$_x$O$_3$\cite{1}.  Correlated phenomena around this compositional quantum critical point was then revealed.  Additionally, an unexpected feature was found in the specific heat data at $x$=0.9.  The feature was believed to reflect the formation of ferromagnetic Ru-rich clusters in the paramagnetic host, however, a full understanding is still lacking \cite{1}. In this short article, we focus on the $x$=0.9 feature and report the results of recent studies to address the issue in details.  

Two new samples were prepared at $x$=0.85 and 0.95 by the same method reported previously \cite{1}.  The magnetic properties were studied in a commercial apparatus (Q.D. MPMS-XL) between 2 K and 390 K below 70 kOe.  Specific-heat measurements were conducted on a small piece of each pellet ($\sim$20-30 mg) in a commercial apparatus (Q.D. PPMS) between 1.8 K and 14.1 K.

\begin{figure}
     \centering
     \includegraphics[width=5cm]{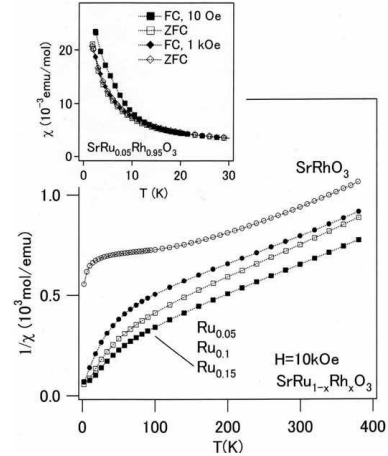}
     \caption{Magnetic susceptibility of SrRu$_{1-x}$Rh$_x$O$_3$ at 10 kOe, measured on cooling.  Small panel: Magnetic susceptibility of SrRu$_{0.05}$Rh$_{0.95}$O$_3$ at low fields, 10 Oe and 1 kOe, on heating to 300 K after cooling the sample without magnetic field (ZFC: zero-field cooling), and then on cooling in the field (FC). } 
\end{figure}  

In Fig.1, magnetic susceptibility data for the newly provided samples ($x$=0.85 and 0.95) are shown with the reported data at $x$=0.9 and 1 \cite{1}. Both samples, as well as the $x$=0.9 sample, clearly show a Curie-Weiss-type character (seen in the $1/\chi$ vs $T$ plots) \cite{1}.  The $x$=1 sample (SrRhO$_3$) is known not to follow this form \cite{1}.  All the plots systematically shift with increasing Ru concentration, indicating the quality of the samples.  What the susceptibility data indicate is not only the sample quality but also the absence of long-ranged magnetic order, consistent with the magnetization data (not shown here, measured at 5 K).  Within the magnetic study above, a relevant feature related to the $x$=0.9 specific heat anomaly was not clearly seen.  

As displayed in the small panel in Fig.1, a weak field dependence of the thermo-magnetic behavior was also studied.  A small thermal hysteresis was seen at 10 Oe for the $x$=0.95 sample and was absent at 1 kOe.  Similar behavior was found for the $x$=0.9 sample \cite{1}, however, it was not observed at $x$=0.85.  The weak field data likely indicate a magnetically glassy freezing such as spin-glass or cluster-glass transitions.  We will mention this below.  

The local crystallographic similarity of both end compounds (SrRhO$_3$ and SrRuO$_3$) vaguely hints at a possible magnetic cluster formation in the solid solution, in which Ru atoms distribute themselves at the Rh sites inhomogeneously.  As is well known, the negative Weiss temperature of itinerant electron systems is not an accurate measure of the strength of magnetic interactions, in contrast to localized electron systems, and long-range ferromagnetic order does appear at higher Ru substitution \cite{1}. The local environment around the possible Ru-rich clusters should therefore, in principle, be ferromagnetic-like.  The $1/\chi$ vs $T$ plots gradually (with Ru substitution) go down from the Curie-Weiss-type linear feature on cooling, indicating a short-range development of the ferromagnetic order, consistent with the cluster model.  The ferromagnetic-like Ru-rich regions in the paramagnetic host behaving like clusters could then be responsible for the observed characteristics in the weak field data.  

\begin{figure}
     \centering
     \includegraphics[width=4.5cm]{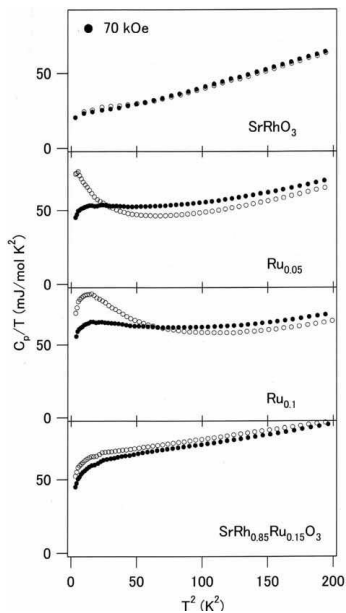}
     \caption{Ru-concentration dependence of the specific heat of the perovskite rhodates measured in a magnetic field of 70 kOe (solid circles) and without the magnetic field (open circles).  Data at $x$=0.9 were taken from our previous report \cite{1}.} 
\end{figure} 

In Fig.2, the Ru concentration dependence and the magnetic field dependence of the specific heat in the low temperature limit are shown.  Two features are seen in the plots upon Ru substitution: enhancement of the coefficient of the linear term, and appearance and disappearance of a small up-turn.  The enhancement might represent contributions from the probable formation of the glassy magnetic components \cite{5}, however the origin of the magnetic up-turn remains uncertain.  The field study revealed the peaks are magnetic origin, however it did not satisfactorily account for the appearance and the disappearance over the substitution range.  Several magnetic models, including a Schottky-type and two empirical terms, were tested on the specific heat data, however none produced a satisfactory fit \cite{1}. Although the electrical resistivity is expected to be influenced by magnetic features, there was no clear evidence in our earlier measurements.  Typically, magnetic features in the transport are subtle, and they may be masked by the polycrystalline nature of the samples \cite{1}.  

In summary, the lightly Ru-doped perovskite rhodates (15 mol$\%$ at most) was reinvestigated by adding new data from magnetic susceptibility and specific heat measurements on samples with $x$ near 0.9.  The major feature of the specific heat data in the vicinity of $x$=0.9 was found to be magnetic in origin and to appear only in a very narrow Ru-concentration range (0.85$<$$x$$\le$0.95 of SrRu$_{1-x}$Rh$_x$O$_3$).  The possible formation of Ru-rich clusters likely plays a significant role in the magnetic behavior.  Further studies on single crystals, as well as theoretical considerations, would be very useful. 

%
%
%
%

%
%
%
%


\end{document}